\documentclass[square]{ws-procs11x85}
\usepackage{balance} 

\newcommand{\met}{\hbox{E\kern-0.5em\lower-0.1ex\hbox{/}}_T}
\newcommand\simlt{\lower.5ex\hbox{$\; \buildrel < \over \sim \;$}}
\newcommand\simgt{\lower.5ex\hbox{$\; \buildrel > \over \sim \;$}}

\begin{document}

\twocolumn[
\title{Collimation and Radiative Deceleration of Jets in TeV AGNs}

\author{A. Levinson and O. Bromberg}

\address{Raymond and Beverly Sackler School of Physics and Astronomy, Tel Aviv University, Tel Aviv 69978, Israel}


\begin{abstract}
We consider some implications of the rapid X-ray and TeV variability observed in M87 and the TeV blazars.
We outline a model for jet focusing and demonstrate that modest radiative cooling can lead to recollimation
of a relativistic jet in a nozzle having a very small cross-sectional radius.  Such a configuration can 
produce rapid variability at large distances from the central engine and may explain recent observations 
of the HST-1 knot in M87.  Possible applications of this model to TeV blazars are discussed.
We also discuss
a scenario for the very rapid TeV flares observed with HESS and MAGIC in some blazars,
that accommodates the relatively small Doppler factors inferred from radio observations.  
\end{abstract}
\vskip12pt  
]

\bodymatter
\section{Introduction}

Recent observations of the VHE emission from compact relativistic systems raise new questions regarding the dynamics 
and dissipation of relativistic jets. Of particular interest is the class of TeV AGNs for which X-ray and TeV 
observations indicate: (i) extremely rapid variability of the VHE emission, in two cases over timescale significantly
shorter than the dynamical time $r_g/c$ of the putative, central black hole;
(ii) substantial differences between the Doppler factor inferred from TeV observations and that 
associated with radio knots;
(iii) rapid variations of the flux emitted from a region located far (at radii $>10^6 r_g$) from the central  
engine.  These observations motivate reconsideration of the standard view, according to which 
dissipation of the bulk energy on sub-parsec scales is accomplished
predominantly through formation of internal shocks in colliding fluid shells.  Below we review the
observational motivation in greater detail and then go on to discuss the implications for
the collimation and dissipation of the outflow.

\subsection{TeV blazars}
Large amplitude variations of the VHE $\gamma$-ray flux on time scales of minuets have been reported for 
Mrk 421, Mrk 501 and PKS 2155-304.   The implied size of the region producing these $\gamma$-ray flares 
(as measured in the Lab frame) is limited to $\Delta r\simlt10^{14}\Gamma \delta t_{\rm var,h}/(1+z)$ cm, 
but is naively expected to exceed $r_g$.  Here $\Gamma$ and $\delta$ are the bulk Lorentz factor and the 
corresponding Doppler factor of the emitting 
matter, respectively, $t_{\rm var,h}$ is the observed variability time in hours, and $z$ is the redshift of the source.   
The extremely short variability reported recently for PKS 2155-304 \cite{ah07} and Mrk 501 \cite{alb07} is therefore
puzzling, as it implies that either, the black hole mass $M_{BH}\simlt10^7 M_{\odot}$, inconsistent with other estimates\cite{woo05},
or that the source imprinting the variability has a characteristic size considerably smaller than the black hole's horizon.
The requirement that the pair production opacity at TeV energies is small implies high Doppler factors, 
$\delta\sim 30 -100$\cite{kra02,lev06}, if the TeV emission originated from jet radii
$r_{\rm em}\sim \Delta r$.  
Such high values are consistent with those obtained from fits of the SED to a homogeneous SSC mode, but are in clear 
disagreement with the much lower values inferred from unification schemes \cite{Urry91,Hardcastle03} and superluminal 
motions on parsec scales \cite{mar99,Jorstad01,Giroletti04}. 

It has been argued that such high values of the Doppler factor may not be required if the 
$\gamma$-ray production zone is located far from the black hole, at radii $r_{\rm em}>>\Delta r$.
This possibility is motivated by recent observations of M87, as described below.
In that case the compactness of the TeV emission zone may be constrained by the variability of the IR flux observed simultaneously
with the TeV flare, allowing low values of $\delta$ in cases where the variability time of the IR emission is much longer than
the duration of the TeV flare.  
However, such a mechanism requires either jet power much larger than the luminosity of the TeV emission measured 
during the flare, $L_{TeV}\sim10^{46}$ erg s$^{-1}$, or focusing of the jet.  

\subsection{M87}
The HST-1 knot in M87 is a stationary radio feature associated with the sub-kpc scale jet.  The knot 
is located at a projected distance of 60 pc ($0.86''$) from the central engine, and is known to be 
a region of violent activity.   Sub-features moving away from the main knot of the HST-1 complex 
at superluminal speeds have been detected recently \cite{cheu07}.  In addition, rapid, large amplitude variations of 
the resolved X-ray emission from HST-1 have been reported, with doubling time $t_{\rm var}$ as short as $0.14$ yrs.
The observed variability limits the linear size of the X-ray source to
$\Delta r \simlt \Gamma\delta t_{\rm var}\sim 0.022\Gamma\delta$ pc, which for reasonable estimates of the 
Doppler factor is three orders of magnitude smaller than the distance between the HST-1 knot and the central black hole.
Based on a claimed correlation between the X-ray and TeV emission it has been proposed that HST-1 may also be the 
region where the TeV emission is produced\cite{cheu07} (but see, e.g., Ref~\refcite{m87} for other explanations).  As mentioned above, this motivated the consideration that the 
TeV emission zone in TeV blazars may also be located far from the black hole.

It has been proposed that HST-1 reflects the location of a recollimation nozzle\cite{staw06}.  
In this picture the superluminal sub-knots that seem to be expelled from the HST-1 complex can be associated with
internal shocks produced by reflection of the recollimation shock at the nozzle.  The rapid variability set a limit
on the cross-sectional radius of the channel at the location of HST-1 that depends on the fraction of jet power 
radiated as X-rays (and TeV emission, if indeed originating from the same location).
Estimates of the jet power in M87 yield $L_j\simgt10^{44}$  erg s$^{-1}$ \cite{bick96,staw06}.
For the observed X-ray\cite{cheu07} and TeV\cite{ah06} luminosities, $L_{\rm TeV}\sim L_{\rm x}\simlt10^{41}$ erg s$^{-1}$, this
implies a rather small radiative efficiency, $\eta\equiv L_{\rm x}/L_j\sim10^{-3}$.  In order to account for the observed luminosity the scale of the fluctuations producing the X-ray flare must satisfy $d\simgt\eta^{1/2} a$, where $a$ denotes the cross-sectional radius of the jet at the dissipation region (see Fig. 1).  The variability time, on the other hand,
constrain the size of these fluctuations to be $d\simlt \Gamma\delta t_{\rm var}$.  Combining the two constraints yields 
$a\simlt\eta^{-1/2}\Gamma\delta t_{\rm var}$.  Associating the apparent speed measured for the superluminal sub-knots in the HST-1 complex with the Lorentz factor of the fluid in the vicinity of the nozzle gives $\Gamma\simgt4$\cite{cheu07}.
Adopting $\Gamma=4$ and viewing angle $\theta_n=30^\circ$, we estimate 
$a\simlt1$ pc for the reported X-ray variability.  
Assuming that the TeV emission is associated with HST-1 would require an even better focusing, $a\simlt0.1$ pc.
At the radius of HST-1 this implies $a/r_{\rm HST1}\simlt 10^{-3}$.

It should be noted that the TeV luminosity observed typically in TeV blazars, $L_{\rm TeV}\sim 10^{44-46}$ 
erg s$^{-1}$, is much larger than X-to-$\gamma$ ray luminosity seen in M87, and is likely to be comparable to the jet power. 
If the rapid TeV variability in the TeV HBLs is to be explained by recollimation of a mildly relativistic jet (with 
fluid Lorentz factor of the order of the apparent superluminal speeds measured in HBLs),
then this would imply far more stringent constraints on the nozzle.  Whether the extreme TeV flares observed in VHE blazars can be accounted for by recollimation shocks at radii $r_{\rm em}>>\Delta r$ remains to be explored.
Recollimation shocks may be an important dissipation channel also in other sources, e.g., GRBs \cite{brom07}.

Below we demonstrate that even modest radiative cooling of the shocked jet layer deflected by the external 
medium can lead to extremely good focusing of the channel.  This has been studied earlier in the non-relativistic regime in 
the context of SS433, e.g., Ref.~\refcite{pe95}

\begin{figure}[t]
\center
\centerline{\psfig{figure=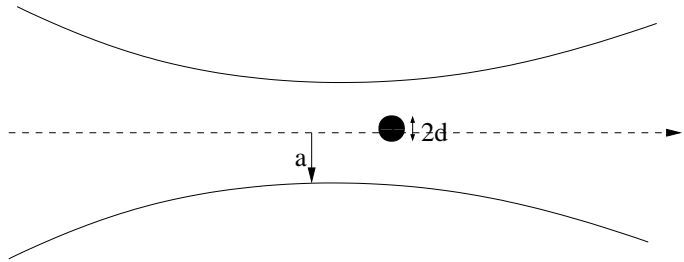,width=9truecm}}
\caption{Sketch of a recollimation nozzle.  The shaded region represents an emitting blob of size $2d$.}
\end{figure}

\begin{figure}[t]
\center
\centerline{\psfig{figure=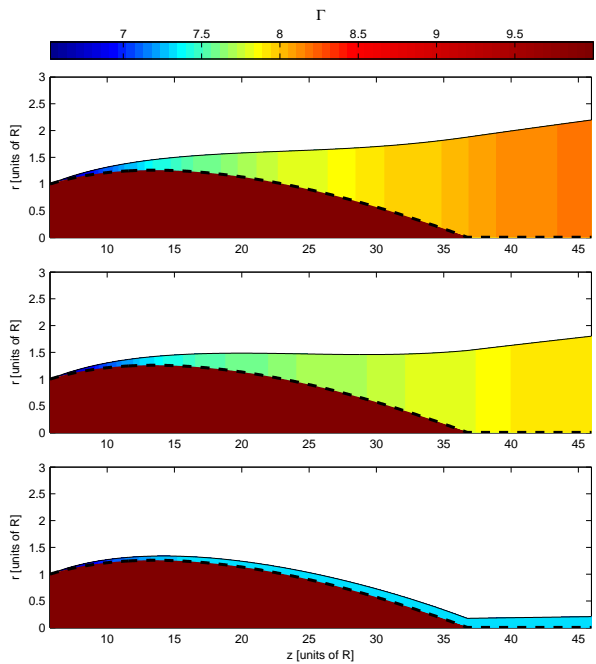,width=9truecm}}
\caption{Lorentz factor of the shocked and unshocked jet in the case of confinement by a static corona with a pressure 
profile $p_{ext}\propto z^{-2}$, for different values of the cooling parameters. 
The ratio of the total luminosity radiated away by the shocked jet layer and total jet power
is $L_c/L_j=0$ (no radiative losses) in the upper panel, 
$L_c/L_j=0.1$ in the middle panel, and $L_c/L_j=0.27$ in the lower panel.
The injected flow in all panels consists of a cold, purely baryonic fluid with
Lorentz factor $\Gamma=10$ at the injection point ($z=0$).  The shock surface is marked by the dashed line}
\end{figure}

\begin{figure}[t]
\center
\centerline{\psfig{figure=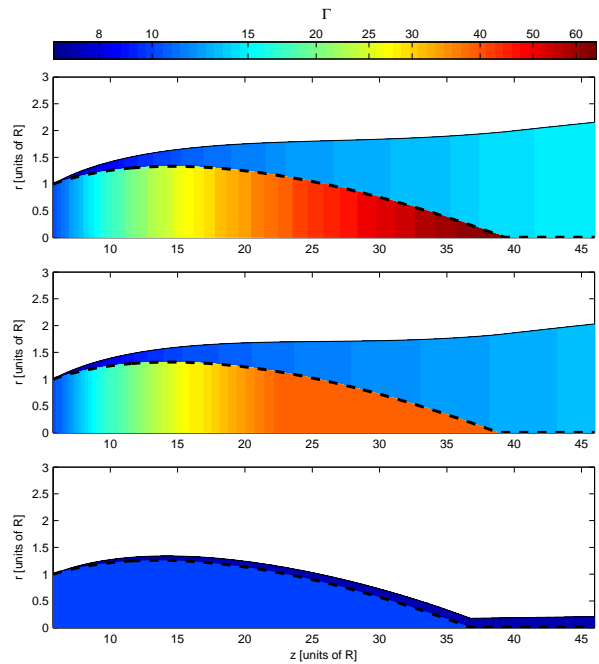,width=9truecm}}
\caption{The effect of jet composition on the collimation process. 
In all cases shown $L_c/L_j=0.27$.  The ratio of total jet energy to rest mass energy is $L_j/\dot{M}_pc^2=1$
in the lower panel (corresponding to the cold, baryonic jet exhibited in Fig. 2), $L_j/\dot{M}_pc^2=4$ in the middle 
panel, and $L_j/\dot{M}_pc^2=10$ in the upper panel. } 
\end{figure}

\section{Focusing of a Radiative Flow by External Medium}
Bromberg and Levinson\cite{brom07} (hereafter BL07) constructed a class of semi-analytical models 
for the confinement and collimation of a relativistic jet by the pressure and inertia of a surrounding medium.
They considered both, confinement by kinetic pressure of a static corona, and confinement by the ram pressure of a supersonic 
wind emanating from a disk surrounding the inner source.  
In general, the collision of the inner jet with the confining medium leads to
the formation of a contact discontinuity across which the total pressure
is continuous, and an oblique shock across which the streamlines of the colliding flow are deflected.
In cases where confinement of the inner flow is accomplished through collision with a supersonic wind
a second shock forms in the exterior wind.  The model outlined in BL07 computes the structure of 
the shocked layers of the deflected inner jet and the exterior wind in the latter case, assuming a 
steady, axisymmetric flow.  In BL07 the focus was on the application to GRBs.  Radiative losses have 
been ignored since the large optical depth of the shocked 
jet layer on scales of interest renders such losses negligibly small.  In blazars, recollimation shocks are expected to 
form above the photosphere.  If a non-negligible fraction of the energy dissipated behind the shock can be tapped for 
the acceleration of electrons to nonthermal energies then the cooling rate may be large enough to affect the structure of the 
recollimation nozzle.  To study this effect we incorporated synchrotron colling into our model.   To be more specific, we 
assume that a fraction $\xi_B$ of the total energy flux is carried by magnetic fields, and that a fraction $\xi_e$ of
the enthalpy behind the shock is injected as a power law distribution of electrons: $dn_e/d\epsilon_e\propto\epsilon_e^{-2}$.

Examples are shown in figures 2 and 3.  In all the examples shown the confining medium is static with a pressure
profile $p_{ext}\propto z^{-2}$.  In Fig. 2 the jet content is dominated by baryons; that is, at the injection point
$L_j\simeq\dot{M}_p\Gamma c^2$ (the energy density of the magnetic field in all examples shown here is taken to
be at most a few percent of the total energy density). The difference between the three cases shown is in the 
synchrotron cooling rate behind 
the oblique shock (which we control by the parameters $\xi_e$ and $\xi_B$).  The ratio of the total luminosity 
$L_c$ radiated away by the shocked jet layer and total jet power $L_j$
injected at $z=0$ in the upper, middle and lower panels is $L_c/L_j=0$ (no radiative losses), 
0.1, and 0.27, respectively.  As seen, in all cases the shock (indicated by the dashed line) converges to the axis
at roughly the same distance from the injection point.  The main effect of the radiative cooling, as seen in Fig. 2, 
is to increase the shock compression ratio, thereby reducing the width of the shocked layer and, as a result, the 
cross-sectional radius of the jet at the recollimation nozzle.  Reflections of the converging shocked layer at the 
symmetry axis should lead to formation of internal shocks in the vicinity of the recollimation nozzle, as seen 
in full 2D simulations.  We propose that the superluminal motions of the sub-features in the HST-1 complex
reported recently are associated with these internal shocks.  A considerable fraction of the remaining jet power 
can dissipate via these internal 
shocks in a region much smaller than what would be expected in the case of a conical jet.  This may lead to 
large amplitude variations over timescales much shorter than the jet radius.  

The effect of cooling is less dramatic in light jets in which the specific enthalpy of the unshocked 
jet is larger than unity, as demonstrated in Fig. 3.  The reason is that in that case the bulk energy 
is dominated by pressure rather than rest mass energy and, therefore, significant fraction of the jet energy must be 
radiated away in order that the shocked jet layer will become under-pressured.  The conditions under which 
effective focusing occurs are under investigation.

\section{Radiative Deceleration of Fluid Shells as the Origin of Rapid Flares in TeV Blazars}

An alternative scenario for the rapid TeV flares has been proposed recently \cite{lev07}, in which flares observed in sources like Mrk 421, Mrk 501 and PKS 2155-304 
are produced by radiative deceleration of fluid shells expelled during violent ejection episodes.
This model cannot naturally account for flares of durations considerably shorter than $r_g/c$, as discussed above. 
The {\em add hoc} assumption made here is that fluid shells accelerate to a Lorentz factor $\Gamma_0>>1$ at some radius $r_d\sim10^2-10^3 r_g$, at which dissipation of their bulk energy occurs.
The dissipation may be accomplished through formation of internal shocks in a hydrodynamic jet or dissipation of magnetic energy in a Poynting flux dominated jet \cite{rom92,lev98}, and it is assumed that a fraction $\xi_e$ of the total
proper jet energy density, $u_j^\prime$, is tapped for acceleration of electrons to a maximum energy $\gamma_{max}m_ec^2$.  
The shocks may also result from a focusing of the outflow, as described in the previous section, which can in principle 
give rise to variability over timescale of the order of $a\Gamma_n^{-2}<r_g/c$, where $a$ again is the cross-sectional radius of the nozzle and $\Gamma_n$ the Lorentz factor of the flow passing the nozzle.  Whether the pressure in the ambient medium is sufficient for significant focusing of the jet and on what scales is yet an open issue.

The fluid equation, 
\begin{equation}
\frac{\partial}{\partial x^\mu}T^{0\mu}= S_c^0,
\label{eq-mot}
\end{equation}
where $S_c^\mu$ accounts for energy losses due to radiative friction, can be solved, assuming that the intensity of ambient radiation depends on radius as $I_s\propto r^{-2}$ and that the proper density and average energy of 
the nonthermal electrons are independent of radius, to yield the asymptotic Lorentz factor\cite{lev07}:

\begin{equation}
\Gamma_\infty=\Gamma_0 \frac{l}{l+r_d},
\end{equation}
where $\Gamma_0=\Gamma(r=r_d)$ is the initial Lorentz factor.  The stopping length $l$ can be expressed in terms 
of the optical depth for $\gamma\gamma$ absorption of a $\gamma$-ray of energy $m_ec^2\epsilon_{\gamma}$ 
by a power law target photon field of the form  $I_s(\epsilon_s)\propto \epsilon_s^{-\alpha}$; $\epsilon_{s,min}<\epsilon_s<\epsilon_{s,max}$, as:
\begin{equation}
\frac{l}{r_d}=\frac{1}{\chi\xi_e\tau_{\gamma\gamma}}\left(\frac{\sigma_{\gamma\gamma}}{\sigma_T}\right)
\left(\frac{\epsilon_\gamma}{\Gamma_0\gamma_{\rm max}}\right)g(\epsilon_\gamma),
\label{stopp2}
\end{equation} 
where $g(\epsilon_{\gamma})\le1$ is some function of energy that depends on the spectral index $\alpha$, and is given explicitly in Ref~\refcite{lev07}, and $\chi=<\gamma^2>/(<\gamma>\gamma_{\rm max})$ is a dimensionless factor that depends on the energy distribution of nonthermal electrons.
For a power law distribution, $dn_e/d\gamma\propto \gamma^{-q}$ with $q\le2$, we have $1>\chi>\>0.1$.

The main conclusion from Eq. (\ref{stopp2}) is that for a reasonably flat distribution of nonthermal electrons, $q\le2$, extension of the distribution to a maximum energy $\gamma_{\rm max}$ at which the pair production optical depth, $\tau(\Gamma_0\gamma_{\rm max})$, is a few is already sufficient to cause appreciable deceleration of the front.
It can be shown \cite{lev07} that for the TeV blazars a background luminosity of $L_s\sim 10^{41}-10^{42}$ erg s$^{-1}$, roughly the luminosity of LLAGN, would lead to a substantial deceleration of the front 
and still be transparent enough to allow the TeV $\gamma$-rays produced by Compton scattering of the background photons
to escape the system.  The ambient radiation field is most likely associated with the nuclear continuum source.  Propagation 
of the $\gamma$-ray flare from low-to-high energies, as reported recently for Mrk 501\cite{alb07}, are naturally expected in this
model, since the $\gamma$-spheric radius increases quite generally with increasing $\gamma$-ray energy\cite{bln95}.
The bulk Lorentz factor of the jet during states of low activity may be appreciably smaller than that of fronts expelled during violent ejection episodes.
 

\section*{Acknowledgments}
This work was supported by an ISF grant for the Israeli Center for High Energy Astrophysics. 
\balance

\end{document}